\documentclass[fleqn,12pt,twoside]{article}
\usepackage{espcrc1,bm}

\usepackage{graphicx}
\usepackage[figuresright]{rotating}

\newcommand{\AmS}{{\protect\the\textfont2
  A\kern-.1667em\lower.5ex\hbox{M}\kern-.125emS}}

\title{$\pi$ and $\rho$ loop corrections to $\omega$ photoproduction in
the resonance region}

\author{Yongseok Oh%
\address[YU]{Institute of Physics and Applied
Physics, Yonsei University, Seoul 120-749, Korea}%
        \thanks{E-mail address: yoh@phya.yonsei.ac.kr}
        and
        T.-S. H. Lee%
\address[Argonne]
{Physics Division, Argonne National Laboratory, Argonne,
Illinois 60439, U.S.A.}%
\thanks{E-mail address: lee@theory.phy.anl.gov
}}
       
\begin{document}

\maketitle

\begin{abstract}
One-loop corrections due to the intermediate $\pi N$ and $\rho N$ states
are studied in $\omega$ photoproduction near threshold.
Our results show that the coupled-channel effects should be taken into
account in extracting reliable nucleon resonance parameters from the
forthcoming vector meson photoproduction data in the resonance region.
\end{abstract}

\section{INTRODUCTION}

The study of photoproduction of light vector mesons such as $\omega$,
$\rho$ and $\phi$ is expected to be useful to resolve the so-called
``missing resonances'' problem and experimental data are now being
accumulated at various experimental facilities
\cite{saphir,clas,graal,leps}.
The extracted $N^*$ parameters can then be used to test existing hadron
models for the baryon resonance structure.
There are some theoretical progress to understand the role of nucleon
resonances in $\omega$ photoproduction at the resonance region
\cite{OTL01,Zhao01}, where the nonresonant amplitudes are computed from
the tree diagrams of the model Lagrangian.
This is, however, obviously not satisfactory by neglecting the hadronic
final state interactions and the coupled-channel effects.
The importance of those effects has been well-known, e.g., in
pion photoproduction, and the first trial
to account for the coupled-channel effects in vector meson
photoproduction has been made in late 1960's \cite{SS68}.
In this work we make an attempt to reinvestigate this problem in
the model of Ref.~\cite{OTL01} with the dynamical formulation developed
in Ref.~\cite{SL96}.

It is the most ideal approach to construct a coupled-channel model by
satisfying the unitarity condition \cite{PM02d}.
However, because of the lack of experimental information which
constrains the transitions between relevant hadronic meson-baryon channels,
we first consider the effects due to intermediate $\pi N$ and $\rho N$
channels only \cite{OL02}.
We also simplify the calculations by only considering the one-loop
corrections which are the leading order terms in a perturbation
expansion of a full coupled-channel formalism.
Thus we can estimate the coupled-channel effects only qualitatively,
but it will be sufficient to test the importance of the
coupled-channel effects in $\omega$ photoproduction near threshold.

\section{MODEL}

\begin{figure}[t]
\centering
\includegraphics[width=0.65\hsize]{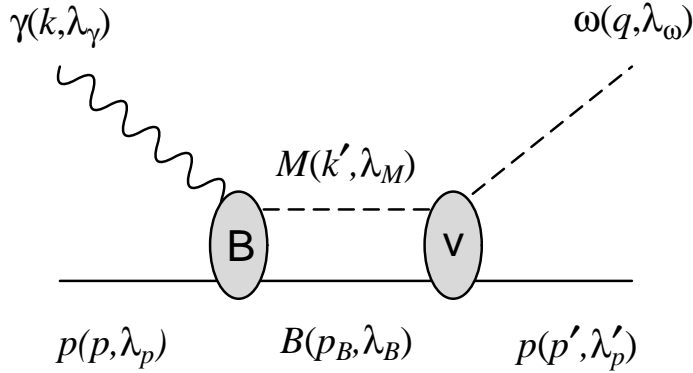}
\caption{Diagrammatic representation of the intermediate meson-baryon
(MB) state in $\omega$ photoproduction.}
\label{fig:diag}
\end{figure}

In the considered energy region, the $\gamma N$ reaction is a
multichannel multiresonance problem.
In this work, following the formalism of Sato and Lee \cite{SL96} we
calculate the one-loop corrections to $\omega$
photoproduction which is represented in Fig.~\ref{fig:diag}.
Then its matrix element in the center of mass frame becomes \cite{OL02}

\begin{eqnarray}
t^{\mbox{\scriptsize one-loop}}_{\gamma N,\omega N}(\bm{k},\bm{q};E)=
\sum_{M=\pi,\rho}\int d\bm{k}^\prime
B_{\gamma N,M N}(\bm{k},\bm{k^\prime};E)G_{MN}(\bm{k}^\prime,E)
v_{M N,\omega N}^{}(\bm{k}^\prime,\bm{q};E),
\label{loop-int}
\end{eqnarray}
where $\bm{k}$ and $\bm{q}$ are the momenta of the incoming photon
and the outgoing vector meson, respectively.
$G_{MN}^{}$ is the propagator for the meson-nucleon system, which reads
\begin{eqnarray}
G_{\pi N}(\bm{q}',E) &=&
\frac{1}{E- E_{N}(\bm{q}^\prime)-E_\pi(\bm{q}^\prime)+i\epsilon},
\nonumber \\
G_{\rho N}(\bm{q}',E) &=&
\frac{1}{E- E_{N}(\bm{q}^\prime)-E_\rho(\bm{q}^\prime)
+i\frac{\Gamma\bm{(}\omega(q',E)\bm{)}}{2}\theta[\omega^2(q',E)-4M^2_\pi]}
\end{eqnarray}
by taking into account the $\rho$ meson width \cite{LEE84}, where
$\theta$ is the step function \cite{OL02}.

We assume that all nonresonant amplitudes $B_{\gamma N,M N}$ (except
pion photoproduction amplitude) and
$v_{M N,\omega N}^{}$ can be calculated from the tree diagrams defined
by the effective Lagrangian \cite{OL02} and the Pomeron exchange.
Therefore the tree diagram model for $\omega$ photoproduction includes
the Pomeron, $\pi$ and $\eta$ exchanges as well as the nucleon
exchanges \cite{FS96,OTL00}.
For the one-loop corrections due to the $\pi N$ channel, we need to know
the amplitudes $B_{\gamma N,\pi N}$ and $v_{\pi N,\omega N}^{}$.
Since there is no tree diagram model for $B_{\gamma N,\pi N}$ in the
considered energy region, we construct the amplitude by subtracting the
resonance contribution from the empirical multipole amplitudes of the SAID
program \cite{SAID}.
Therefore, $B_{\gamma N,\pi N}$ for $\gamma p \to \pi^0 p$ and $\gamma p
\to \pi^+ n$ are calculated as in
Ref. \cite{DGL02} except that we use the resonance parameters from
PDG \cite{PDG00}.
For the $\pi N \to \omega N$ amplitude, we consider the $\rho$ and
$b_1(1235)$ exchanges in $t$-channel and the nucleon exchange in $s$- and
$u$-channel.
We found that the contribution of the $b_1$-meson exchange is suppressed in the
considered energy region and our results are consistent with the
nonresonant amplitudes of Ref.~\cite{PM01a}.

Next we consider the one-loop corrections due to the $\rho N$ channel.
The very limited data show that $\rho^\pm N$ and $\rho \Delta$
photoproduction processes are much weaker than $\rho^0$ photoproduction.
Therefore, we keep $\rho^0 p$ only in the loop calculation
(\ref{loop-int}).
The $\rho^0$ photoproduction amplitudes are constructed by the Pomeron,
$\sigma$, $\pi$ and $\eta$ exchanges together with the nucleon exchange
\cite{OL02,FS96,OTL00}.
The $\rho p \to \omega p$ amplitude includes the $\pi$ and nucleon
exchanges.
This amplitude is related with the tree diagrams of $\omega$
photoproduction in the vector dominance model, except that the
$\rho p \to \omega p$ amplitude does not allow the Pomeron and $\eta$
exchanges because of their quantum numbers.
Form factors are included for each process and fitted by available experimental
data.
The details on the amplitudes and comparison with the data can be found
in Refs.~\cite{OL02,OTL00}.

\section{RESULTS AND DISCUSSION}

\begin{figure}[t]
\includegraphics[angle=-90,width=0.63\hsize]{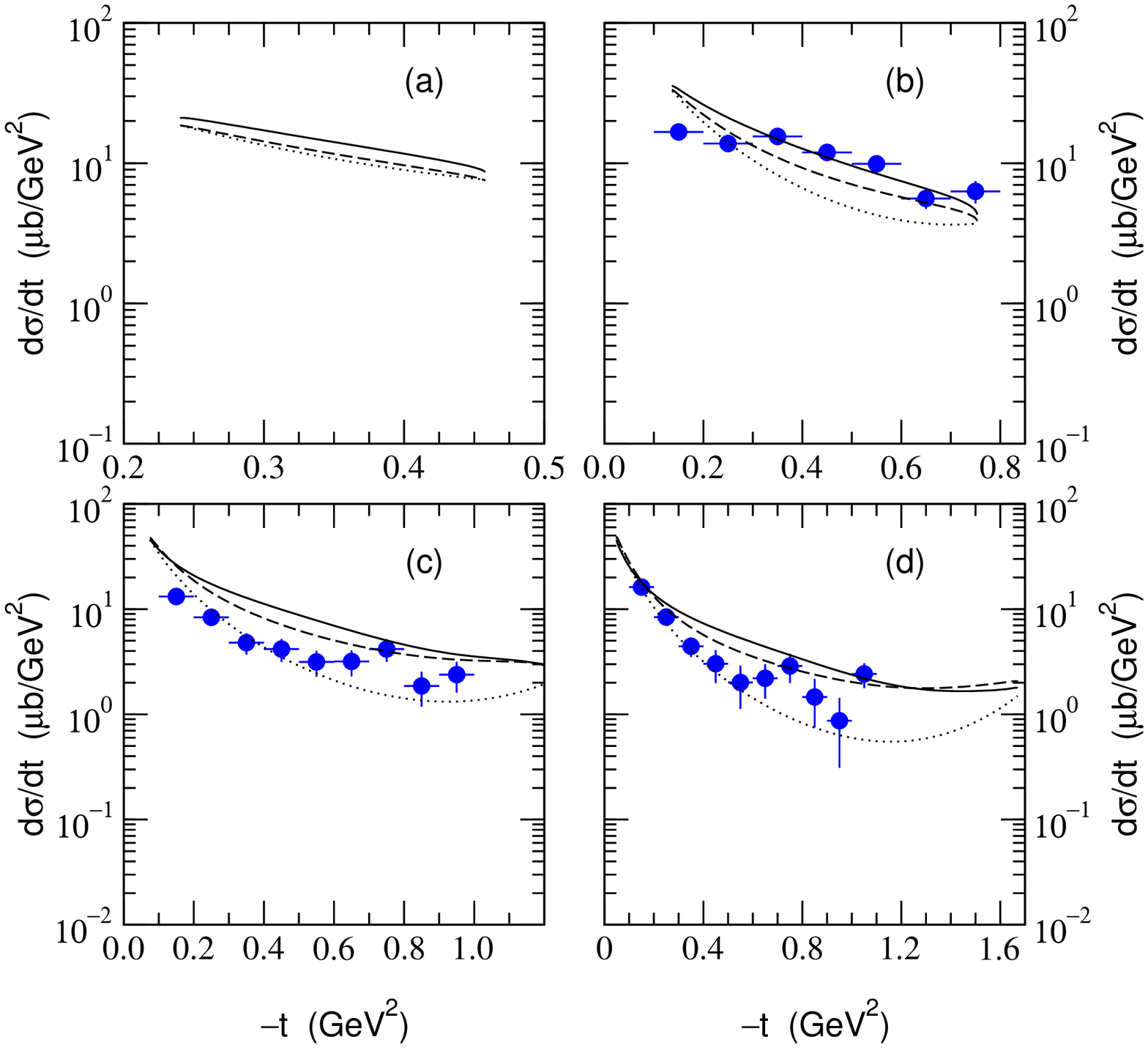}
\hspace{-2.5cm}
\includegraphics[angle=-90,width=0.63\hsize]{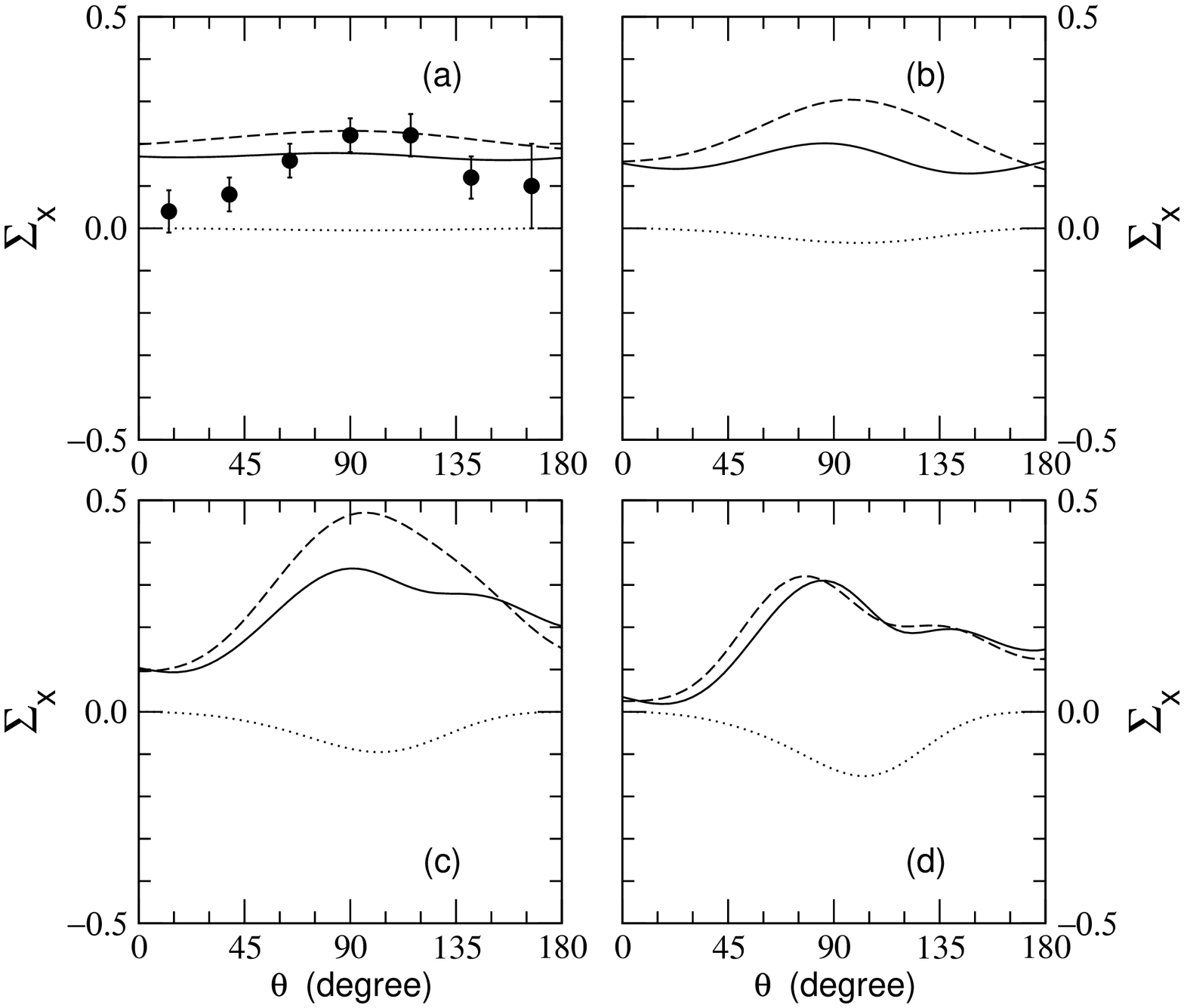}
\caption{Differential cross section (left panel) and single photon
asymmetry $\Sigma_x$ (right panel) for $\gamma p \to \omega p$ 
at $E_\gamma = $ (a) 1.125 GeV, (b) 1.23 GeV, (c) 1.45 GeV and (d) 1.68
GeV. The dotted lines are from the tree diagrams only and the dashed
lines include the tree diagrams and the intermediate $\pi N$ channel.
The solid lines are the full calculations including the $\rho N$ channel
in addition. The experimental data are from Refs.~\cite{saphir,graal}.}
\label{fig:result}
\end{figure}

With the amplitudes constructed above, we can now investigate the
one-loop corrections due to the intermediate $\pi N$ and $\rho N$
channels in $\omega$ photoproduction.
In the left panel of Fig.~\ref{fig:result} we show the results of
differential cross sections of $\omega$ photoproduction.
This shows that the role of the intermediate meson-baryon channels is
very important to understand the production process near threshold.
Although the magnitudes of the one-loop corrections are small compared
with the tree diagram ones, their interference makes the contribution of
the intermediate meson-baryon channels nontrivial.
However, because of the uncertainties and the approximations made in the
amplitudes, the results should be regarded as a qualitative indication
for the role of the coupled-channel effects.

In addition to the differential cross sections, various spin polarization
asymmetries have been suggested to identify the role of nucleon resonances
\cite{OTL01,Zhao01}.
It is thus legitimate to test the coupled-channel effects within such
asymmetries.
Given in the right panel of Fig.~\ref{fig:result} are the predictions
for the single photon asymmetry \cite{TOYM98}.
This result confirms that the coupled-channel effects are very important
in polarization asymmetries.

In summary, we investigate the one-loop corrections to $\omega$
photoproduction near threshold as a step toward developing a
coupled-channel model for vector meson photoproduction.
This calculation was performed by assuming that all relevant nonresonant
amplitudes can be calculated from tree diagrams of effective
Lagrangians.
Together with the fact that the experimental information is not
sufficient to constrain the relevant transition amplitudes, the computed
one-loop corrections are just the leading terms of a perturbative
expansion for a full coupled-channel model.
Therefore our results should be taken as a qualitative indication of the
importance of the coupled-channel effects.
However the results show that the coupled-channel effects should be
carefully taken into account in extracting the
resonance parameters from the forthcoming experimental data, especially
the data of polarization observables.

\section*{ACKNOWLEDGMENTS}
The work of Y.O. was supported by Korea Research Foundation Grant
(KRF-2002-015-CP0074). T.-S.H.L. was supported by U.S. DOE Nuclear
Physics Division Contract No. W-31-109-ENG-38.

\end{document}